# Interpreting the semi-classical stress-energy tensor in a Schwarzschild background, implications for the information paradox

James M. Bardeen

*Physics Department, Box 1560, University of Washington*
*Seattle, Washington 98195-1560. USA*
*bardeen@uw.edu*

**Abstract**

Numerical results for the semi-classical stress-energy tensor outside the horizon of a Schwarzschild black hole obtained in the 1980's and 1990's are re-examined in order to better understand the origin of Hawking radiation and the implications for the black hole information paradox. Polynomial fits to the numerical results for the 4D transverse stress are obtained for conformally-coupled spin 0 and spin 1 fields in the Hartle-Hawking and Unruh states and. Analysis of the spin 0 Unruh state results clearly shows that the origin of the Hawking radiation is not pair creation or tunneling very close to the black hole horizon, but rather is a nonlocal process extending beyond $r = 3M$. Arguments are presented that the black hole information paradox cannot plausibly be addressed by processes occurring on or very close to the horizon of a large black hole whose geometry is close to Schwarzschild.

## I. INTRODUCTION

The original derivation of Hawking radiation from black holes[1] and the prediction of its essentially thermal character at the Hawking temperature

$$T_H = \frac{\kappa m_p^2}{2\pi} \quad (G = c = 1, \hbar = m_p^2), \tag{1.1}$$

with $\kappa$ the surface gravity of the horizon, was based on semi-classical effective field theory. For a Schwarzschild black hole of mass $M$, $\kappa = 1/4M$. I do not set the Planck mass $m_p$ equal to one, in order to emphasize the smallness of quantum corrections, of order $m_p^2 / M^2 < 10^{-76}$, for $M > 1 M_\odot$, for a large astrophysical black hole. In the semi-classical approximation quantum fields propagate in a classical background spacetime, a solution of the classical Einstein equations. The Hawking luminosity for a massless field of spin $s$ has been parameterized as

$$L_H = \frac{4\pi}{245760\pi^2} \frac{m_p^2}{M^2} k_s = 4\pi M^2 \sigma T_H^4 k_s, \tag{1.2}$$

where $\sigma = \pi^2 / (60 m_p^6)$ is the Stefan-Boltzmann constant. For a solar mass or larger black hole only spin 1 photons and spin 2 gravitons are expected to contribute, for which Page[2] found $k_1 = 6.4928$ and $k_2 = 0.7404$, respectively. A hypothetical conformally-coupled massless scalar field has $k_0 = 14.36$ (Elster[3]; Taylor, Chambers, and Hiscock[4]). Later calculations determined the expectation value of the complete renormalized semi-classical stress-energy tensor (SCSET) for conformally-coupled spin 0 and spin 1 fields outside the horizon, $r > 2M$, using the point-splitting renormalization procedure of Christensen[5] and applied to black holes by Christensen and Fulling (CF)[6]. The standard assumptions are that the SCSET is time-independent (neglecting the back-reaction on the geometry) and must satisfy local energy-momentum conservation. The trace of the SCSET is zero classically, but the renormalization breaks the conformal invariance of the fields to produce an anomalous trace depending only on the local curvature of the background spacetime and the spin of the field[7].

    The calculations require specifying the quantum state. While historically this has been called a choice of vacuum, this is an unfortunate misnomer, since what is "vacuum" is ill-defined in an inhomogeneous curved spacetime for modes with wavelengths the order of the curvature scale. An unambiguous definition of particles as excitations of a "vacuum" is only possible in the approximately Minkowskian spacetime well before formation of the black hole or in the asymptotically flat region near future null infinity. The two states usually considered are the Hartle-Hawking (HH) state[8], the thermal state for an eternal black hole in equilibrium with an external heat bath at the Hawking temperature, and the Unruh state[9] appropriate for a black hole formed by gravitational collapse, once transients associated with formation of the black hole have decayed, The Hartle-Hawking state is static, with no net energy flux as seen by a static observer anywhere outside the future and past horizons. The Unruh state has no incoming radiation at past null infinity and outgoing radiation at future null infinity. Both are regular in a freely falling frame at the future horizon.

    Numerical results for the SCSET of a conformally-coupled spin 0 field in the HH state in a Schwarzschild background were obtained by Howard and Candelas[10] (HC), and later with improved accuracy by Anderson, Hiscock and Samuel (AHS)[11]. Elster[12] attempted an extension to the spin 1 HH state, but made an error in his application of point-splitting renormalization. Corrected spin 1 HH results were published by Jensen and Ottewill (JO)[13]. Elster's calculation[3] of the difference between the Unruh and HH SCSETs for a spin 0 field was more successful. Jensen, McLaughlin, and Ottewill[14] (JMO) improved on his results and extended them to the spin 1 case. It was only necessary to calculate the transverse stress component $<T_\theta^\theta>$ from scratch, since the conservation laws, supplemented by boundary conditions appropriate to the quantum state, and the trace anomaly determine the remaining components of the SCSET for a spherically symmetric black hole. Most of these papers presented the results only in rather crude graphical form. Fortunately, Visser[15] published tabular data from the JO and JMO calculations for the spin 0 HH and Unruh states and has preserved computer files of the JO and JMO results for spin 1, which he was kind enough to share.[16]

    Analytic expressions fitted to the numerical results over a finite range of radius allow extrapolation to infinite radius and facilitate their interpretation. Visser[15] found a fit to the transverse stress of a conformally-coupled spin 0 field in

the Unruh state as a polynomial in $2M/r$. However, his fit is not quite consistent with the value of the spin 0 Hawking luminosity quoted above. In Part II of the paper I consider carefully the framework for such fits, and using the data tables from Visser obtain precise polynomial fits to the transverse stress for both spin 0 and spin 1 HH states and for the spin 0 Unruh state. However, the range and accuracy of the available spin 1 Unruh state data are not sufficient for an unambiguous extrapolation to large radii. Requiring consistency with the Page result for the Hawking luminosity and using a fitting procedure which is not sensitive to errors in the data at larger values of the radius gives what I consider a plausible result, but more accurate calculations extending to larger radii are required to firm up the spin 1 results.

The physical interpretation of the SCSET is taken up in Part III, with the focus on the Unruh state. The Unruh state energy flux in static frames is an outward flow of positive energy at large $r$, but at the horizon must represent an inward flow of negative energy if the SCSET is to be nonsingular in a falling frame. How and where the transition between these limits occurs is an indication of where the Hawking radiation is being generated. I argue that the SCSET is inconsistent with the Hawking radiation being generated very close to the horizon. The conversion of vacuum fluctuations into Hawking radiation is something that happens nonlocally in the general vicinity of $r = 3M$, not the horizon at $r = 2M$.

In Part IV, I discuss the first-order back-reaction on the geometry associated with the SCSET on and outside the horizon. The accurate results for the SCSET confirm the qualitative picture of black hole evaporation widely accepted in the literature, and the conclusion of Bardeen[17] that the metric remains Schwarzschild to a very good approximation outside the horizon as the black hole evaporates.

Part V considers of the implications for the black hole information paradox[18]. I argue that storing a significant amount of quantum information in a non-degenerate "stretched horizon"[19] or in a "thermal atmosphere" is not possible as long as the backreaction is small and the semi-classical approximation is valid. The "stretched" horizon is just a way of dealing with *external* perturbations of a classical black hole. The "thermal atmosphere" is not a property of the black hole, it is a property of an accelerating particle detector, whether near a black hole horizon or in ordinary Minkowski spacetime. Trapped quantum information ends up in the deep interior of a Schwarzschild black hole on a dynamical time scale, and is not available to resolve the information paradox, assuming it propagates causally in a Schwarzschild background geometry.

The black hole information paradox is put most starkly when the quantum fields are initially in a pure state, with zero von Neumann entropy. The emission of Hawking radiation results in increasing entanglement between the fields in the exterior of the black hole and those in the interior. In the semi-classical theory there seems to be no limit to the emission of Hawking radiation and growth of the entanglement entropy until the black hole has shrunk down to the Planck scale. This seems hard to square with the thermodynamic Bekenstein-Hawking entropy proportional to the area of the black hole horizon, and the basic principle of unitarity in quantum mechanics that for any complete quantum system pure states evolve into pure states.

The classical evolution of the black hole is governed by theorems that rely on energy conditions easily violated locally in quantum field theory. In particular, the classical focusing theorem for hypersurface-orthogonal null geodesic congruences, together with the null energy condition $T_{\alpha\beta}k^\alpha k^\beta \geq 0$, is the basis for proving the existence of an event horizon permanently shielding the interior of the black hole from observations in the exterior[20]. Proposed quantum null energy conditions are tested against the SCSET. The Quantum Focusing Conjecture (QFC) of Bousso, et al[21] can be used to derive some of these conditions and could a basis for a generally applicable theory of large quantum black holes. It implies a *quantum* singularity theorem similar to the classical Penrose singularity theorem[22]. However, the QFC may not be valid in the context of quantum gravity, and in any case is limited to a semi-classical context.

## II. THE SEMI-CLASSICAL STRESS-ENERGY TENSOR OUTSIDE THE SCHWARZSCHILD HORIZON

The SCSET is the expectation value of the renormalized energy-momentum tensor of quantum fields calculated to first-order in $\hbar$ on a fixed classical background geometry, taken here to be the spherically symmetric Schwarzschild geometry, with the metric
$$ds^2 = -(1-2M/r)dt^2 + (1-2M/r)^{-1}dr^2 + r^2(d\theta^2 + \sin^2\theta\, d\varphi^2). \qquad (2.1)$$
At $r > 2M$ it is convenient to work with the physical components of the SCSET as projected onto the orthonormal frames of static observers, uniformly accelerating observers whose world lines are at constant Schwarzschild radius r. However, it is important to remember that static observers are unphysical in the limit $r \to 2M$, where their proper acceleration becomes infinite. The only physical frames on the Schwarzschild future horizon are "falling" frames. The global geometry is not static.

For the quantum states considered here, the SCSET is spherically symmetric, with four independent components, an energy density $E = -T^t_t$, an energy flux/momentum density $F = -T^r_t/(1-2M/r)$, a radial stress $P_r = T^r_r$, and a transverse stress $P_t = T^\theta_\theta = T^\varphi_\varphi$, all as defined in the frame of a static observer. Outside the horizon any classical disturbances not protected by global conservation laws dissipate by a combination of radiation out to future null infinity and inward across the horizon to the black hole interior, unless associated with persistent external sources. After transient behavior associated with black hole formation, on a time scale of order several times $M$, the expectation value of the energy-momentum tensor of a quantum field is assumed to become stationary to first-order in $\hbar$.

As an expectation value, the SCSET should be considered an average over times very long compared with $M$, but very short compared with the evaporation time of order $M^3/m_p^2$. As a tensor, the components in a different local orthonormal frames at a given point in spacetime depend only on the relative 4-velocity of the frames. The SCSET says nothing about what a local particle detector measures, since that is very sensitive to the *acceleration* of the detector, even in Minkowski spacetime, as noted by Unruh and Wald[23]. The SCSET must be conserved to be a source in the classical Einstein equations.

With time derivatives set to zero, the energy conservation equation is

$$r^2 T^{\beta}_{t;\beta} = \partial_r \left[ r^2 \left(1 - \frac{2M}{r}\right) F \right] = 0 \tag{2.2}$$

and momentum conservation is

$$r T^{\beta}_{r;\beta} = (E + P_r) \frac{M/r}{1 - 2M/r} + \frac{1}{r} \partial_r (r^2 P_r) - 2 P_t = 0. \tag{2.3}$$

Since the Hawking luminosity $L_H = \lim_{r \to \infty} (4\pi r^2 F)$, we see from Eqs. (1.2) and (2.2) that the spin $s$ contribution to the net energy flux at finite $r$ is

$$F = \frac{k_s}{245760} \frac{m_p^2}{M^2 r^2} \frac{1}{(1 - 2M/r)} = k_s \sigma T_H^4 \frac{M^2}{r^2} \frac{1}{(1 - 2M/r)}. \tag{2.4}$$

For conformally coupled quantum fields, the only contribution to the trace of the SCSET is the trace anomaly (also called the conformal or Weyl anomaly). In the Schwarzschild background, the Ricci tensor $R_{\alpha\beta}$ and the scalar curvature $R$ vanish, and for spin $s$

$$T^{\alpha}_{\alpha} = q_s \frac{m_p^2}{2880 \pi^2} \left( R_{\alpha\beta\gamma\delta} R^{\alpha\beta\gamma\delta} = 48 \frac{M^2}{r^6} \right). \tag{2.5}$$

The coefficient is $q_0 = 1$ for a conformally-coupled massless spin 0 field, $q_{1/2} = 7/2$ for a massless spin 1/2 (Dirac) field, $q_1 = -13$ for the spin 1 electromagnetic field, and $q_2 = 212$ for a massless spin 2 field. Then, with $x \equiv 2M/r$,

$$T^{\alpha}_{\alpha} = -E + P_r + 2P_t = 64 q_s \sigma T_H^4 x^6. \tag{2.6}$$

The numerical calculations make use of the constraints imposed by the conservation of the SCSET and the knowledge of its trace from the trace anomaly. Only the transverse stress needs to be calculated from scratch, since the momentum conservation equation, Eq. (2.3), can be integrated to find the radial stress.

From Eq. (2.4) the static frame energy flux, if nonzero, is infinite at the horizon. This is perfectly consistent with a SCSET that is nonsingular in a falling frame, as normally assumed, since the static frame at the horizon is moving outward at the speed of light relative to any frame freely falling from a nonzero distance outside the horizon. Any ingoing energy in the free-fall frame is infinitely blueshifted in the Lorentz transformation to the static frame, just as any outgoing energy is infinitely redshifted. An energy flux finite in a falling frame at the horizon, together with a positive Hawking luminosity, means that in the static frame the SCSET is dominated by inflow of negative energy near the horizon. The infinite boost for ingoing energy produces infinite energy density and radial stress, but leaves the transverse stress unaffected. In order to isolate the singular behavior on the horizon, define an "ingoing" part of the SCSET, $\langle T^{\alpha}_{\beta} \rangle^{in}$, which is defined in the static frame by

$$E^{in} = P_r^{in} = -F^{in} = -F = -\frac{1}{4}k_s\sigma T_H^4 \frac{x^2}{1-x}, \quad P_t^{in} = 0. \tag{2.7}$$

By itself, this satisfies momentum conservation, as can be verified from Eq. (2.3), and is traceless.

I call the remainder of the SCSET the "regular" part, $\left(T_\mu^\nu\right)^{reg}$, with static frame components $E^{reg}$, $P_r^{reg}$, and $P_t$. It seems natural to assume that the "regular" stresses can be approximated by polynomials in $x$, with coefficients constrained by the momentum conservation equation (2.3). Terms of the form $x^n \log(x)$ are also perfectly regular at the horizon, but they complicate the asymptotic behavior at infinity, and don't seem to improve fits to the numerical data in a useful way. Therefore, assume the forms

$$P_t = h_s P_0 \sum_{n=0}^{N} t_n x^n, \quad P_r^{reg} = h_s P_0 \sum_{n=0}^{N} r_n x^n. \tag{2.8}$$

The quantity $P_0 \equiv 2\sigma T_H^4 / 3$ is the thermal pressure per helicity state for a massless field at the Hawking temperature, and $h_s$ is the number of helicity states for spin $s$, with $h_s = 1$ for a spin 0 field and $h_s = 2$ for a spin 1 field. The regular part of the energy density is then $E^{reg} = P_r^{reg} + 2P_t^{reg} - T_\alpha^\alpha$. The order $N$ of the polynomials must be at least 6 in order to accommodate the trace anomaly. The trace anomaly is $T_\alpha^\alpha = 96 q_s x^6 P_0$, or $96 x^6 P_0$ for spin 0 and $-1248 x^6 P_0$ for spin 1.

The formal solution of the momentum conservation equation (2.3) governing the regular part of the SCSET is

$$P_r^{reg} = \frac{x^2}{1-x} \int_1^x \left[ \frac{(3y-2)}{y^3} P_t(y) - \frac{1}{2y^2} T_\mu^\mu(y) \right] dy. \tag{2.9}$$

A polynomial for $P_t$ is consistent with a polynomial for $P_r^{reg}$ if and only if $t_2 = 3t_1/2$, as is true for both the HH and Unruh states. For the Unruh state the asymptotic SCSET is dominated by the radial outflow of the Hawking radiation, since there is no incoming radiation. Therefore, $t_0 = t_1 = 0$ and

$$E \cong P_r \cong F \to \frac{1}{4} k_s \sigma T_H^4 x^2 = \frac{3}{8} k_s P_0 x^2. \tag{2.10}$$

This requires $P_r^{reg} \to 2F$ in order to compensate for $P_r^{in} = -F$, and thus

$$r_2 = \frac{3k_s}{4h_s}, \tag{2.11}$$

This is a condition on the value of the integral in Eq. (2.9) for $x=0$, and imposes a constraint on the coefficients in the expression for $P_t$. While in principle one could determine $r_2$ from an extremely accurate fit to the numerical data for $P_t$, this is not advisable, since the numerical data only extend over a finite range of $r$ and there are likely to be significant errors in extrapolation to $x \ll 1$. The direct calculations of the Hawking luminosity are much more accurate.

Evaluating the integral in Eq. (2.9) for $x=0$ gives

$$r_2 = \sum_{n=0}^{N} \frac{4-n}{(n-1)(n-2)} t_n - \frac{48 q_s}{5 h_s}. \quad (2.12)$$

This can be considered a constraint on the value of $t_N$. For instance, if $N = 6$,

$$t_6 = -10 r_2 - 45 t_1 + 5 t_3 - \frac{5}{6} t_5 + \frac{96 q_s}{h_s}, \quad (2.13)$$

and for $N = 7$,

$$t_7 = -10 r_2 - 45 t_1 + 5 t_3 - \frac{5}{6} t_5 - t_6 + \frac{96 q_s}{h_s}. \quad (2.14)$$

The remaining $r_n$ can be found from the recursion relations

$$(n-2) r_n = (n-2) r_{n-1} - 2 t_n + 3 t_{n-1} - \frac{48 q_s}{h_s} \delta_{n7}. \quad (2.15)$$

The regularity condition $\left(E^{\text{reg}} + P_r^{\text{reg}}\right)_{x=1} = 0$ is satisfied identically.

The asymptotic conditions on the SCSET appropriate to the HH state, for which $\left\langle T_\alpha^\beta \right\rangle^{\text{in}} \equiv 0$, correspond to equilibrium of a thermal gas in the static frame,

$$P_r = P_t = h_s P_0 (1-x)^{-2} = h_s P_0 \left(1 + 2x + 3x^2 + \ldots\right). \quad (2.16)$$

This requires $t_0 = 1$, $t_1 = 2$ $(\Rightarrow t_2 = 3)$, and $r_2 = 3$. Deviations from the thermal gas expansion are expected at order $x^3$, due to the geodesic deviation associated with the background curvature. There was some confusion about this point in the early literature, in which there seemed to be an expectation that the first deviations should go as the square of the curvature, at order $x^6$ (the "strong thermal hypothesis").

For the Unruh state, assuming that the asymptotic SCSET corresponds to classical radiation emitted from the black hole shadow, one would expect that the first nonzero $t_n$ should be $t_4$. However, a good fit to the numerical results requires $t_3 \neq 0$ for both spins.

If the percentage numerical errors in calculating $P_t$ were uniform, it would seem best to determine the $t_n$ by a least squares to fit to $x^{-3} P_t$, in view of the way $P_t$ enters the integral for $P_r^{\text{reg}}$ in Eq. (2.9). For the spin 1 Unruh state, with the magnitude of $P_t$ falling by a large factor in going from the horizon to the outer limit of the numerical results at $x = 0.4$, this assumption does not seem to be justified.

### a) Spin 0 Hartle-Hawking state

The Page approximation[24] to the total spin 0 HH transverse stress is

$$P_t = P_0 \left(1 + 2x + 3x^2 + 4x^3 + 5x^4 + 6x^5 - 9x^6\right). \quad (2.17)$$

While often cited as remarkably accurate, this differs from the numerical results of Howard and Candelas[9] and of Anderson, et al[10] by about 50% around $x \sim 0.8$, and by about 17% at $x = 1$. The se errors are indeed small compared with the scale set by the trace anomaly. The approximation could easily be improved to agree with the Candelas result for $P_t(1)$ by setting $t_4 = 3.27$. A change in $t_4$ does not induce any compensating change in the other $t_n$.

The Page approximation satisfies the "strong thermal" hypothesis. However, allowing a departure from the thermal gas expansion at order $x^3$ results in a remarkably good $N = 6$ polynomial fit to the AHS[11] data as quoted by Visser[15], with two free parameters,

$$P_t = P_0\left(1 + 2x + 3x^2 + 3.650x^3 + 14.398x^4 - 48.170x^5 + 34.392x^6\right). \quad (2.18)$$

The maximum errors are 0.03%, with $\chi^2 = 1.35 \times 10^{-5}$. The formal uncertainty in the fitted value of $t_3$ is only 0.005. That such a good fit is possible with two free parameters is testimony both to the accuracy of the AHS numerical results and to how closely this fit must match the exact result. Allowing one more free parameter, in a $N = 7$ fit, reduces the $\chi^2$ only slightly, to $\chi^2 = 1.23 \times 10^{-5}$, and the uncertainty in the coefficients becomes several times larger, with a value of $t_7$ consistent with zero.

### b) Spin 0 Unruh state

The polynomial fit found by Visser[15] to the spin 0 Unruh state transverse stress based on the AHS results for the HH state and the JMO results for the Unruh - HH difference, starting at $x^4$ as argued by CF[6], is

$$P_t = P_0\left(26.562x^4 - 59.0214x^5 + 38.0268x^6\right). \quad (2.19)$$

The maximum residual is about 0.7% around $x = 0.5$. This fit made no use of the Hawking luminosity, and the $r_2$ determined just from the transverse stress fit corresponds to $k_0 = 14.26$. However, direct calculation of the Hawking luminosity is simpler and should be more accurate. The spin 0 luminosity found originally by Elster[3] is $L_H = 7.44 \times 10^{-5}\left(m_p / M\right)^2$, which implies $k_0 = 14.36$ and $r_2 = 10.77$. This value was confirmed to the full three significant figures by independent calculations of Simkins[25] and of Taylor, Chambers and Hiscock[26]. While the difference seems small, it is significant because of the factor of 10 multiplying $r_2$ in Eq. (2.13).

Adopting $r_2 = 10.77$, the 4-6 fit to $x^{-3}P_t$ gives $t_4 = 27.953$, $t_5 = -62.764$, with a rather poor $\chi^2 = 0.0602$. Allowing a nonzero $t_3$ results in a much better fit,

$$P_t = P_0\left(0.2524x^3 + 25.5439x^4 - 57.6663x^5 + 37.6172x^6\right), \quad (2.20)$$

with $\chi^2 = 0.000903$. The formal uncertainties in the free parameters $t_3, t_4, t_5$ are $\pm 0.0065$, $\pm 0.063$, $\pm 0.14$ respectively, all quite small compared with their values. The residuals of the two fits are compared in Fig. 1. Allowing $r_2$ to be an additional free parameter in the 3-6 fit gives $r_2 = 10.87 \pm 0.29$ and does not substantially reduce the $\chi^2$. For both fits I slightly altered the values at the data points $r/2M = 1.7$ and $r/2M = 2.1$ to make the residuals there more consistent with those at neighboring points.

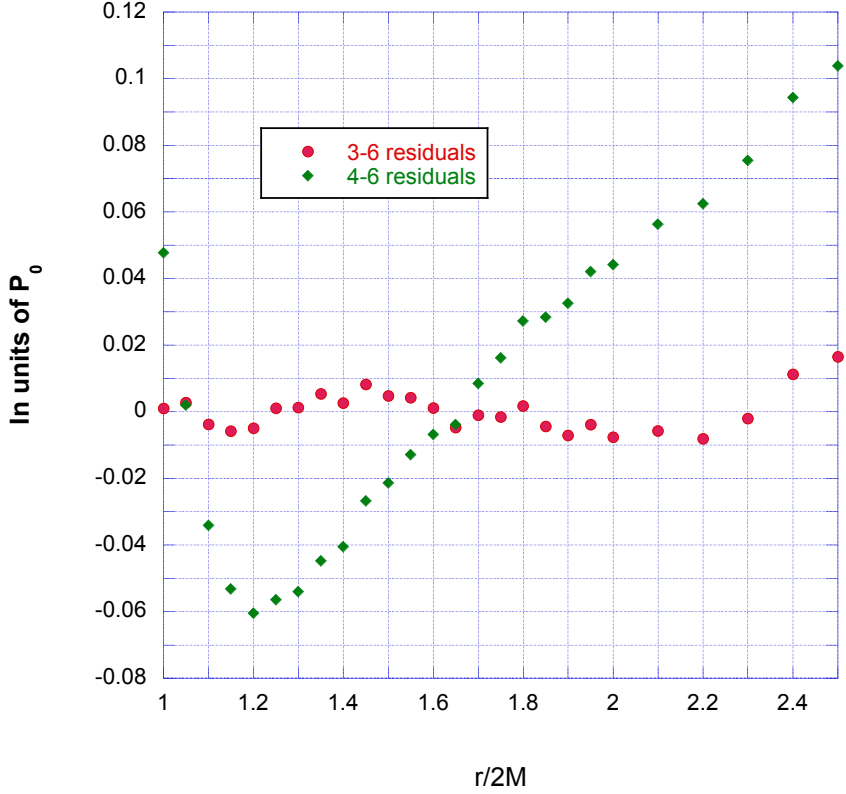

Figure. 1. Comparison of residuals from fitting the spin 0 Unruh state $x^{-3}P_t$ from the data given in Table 2 of Visser[14] to $n = 3-6$ and $n = 4-6$ polynomials, keeping a fixed $r_2$.

From Eqs. (2.20) and (2.15), the "regular" part of the radial stress for the 3-6 fit is

$$P_r^{\text{reg}} = P_0 \left(10.77 x^2 + 10.2652 x^3 - 14.9001 x^4 + 49.0880 x^5 - 12.9703 x^6\right). \quad (2.21)$$

All of the components of the SCSET for spin 0 Unruh state are shown together in Fig. 2. Within the range of the numerical data, $x \geq 0.4$, the errors from the 3-6 fit are much less than the width of the lines. The signs of what is plotted have been chosen to make it clear that $E^{\text{reg}} + P_r^{\text{reg}}$ is zero at the horizon. Also, note the approach toward domination by a radial outflow of Hawking radiation as $x \to 0$, with $x^{-2} E^{\text{reg}} \cong x^{-2} P_r^{\text{reg}} \cong 2 x^{-2} F$.

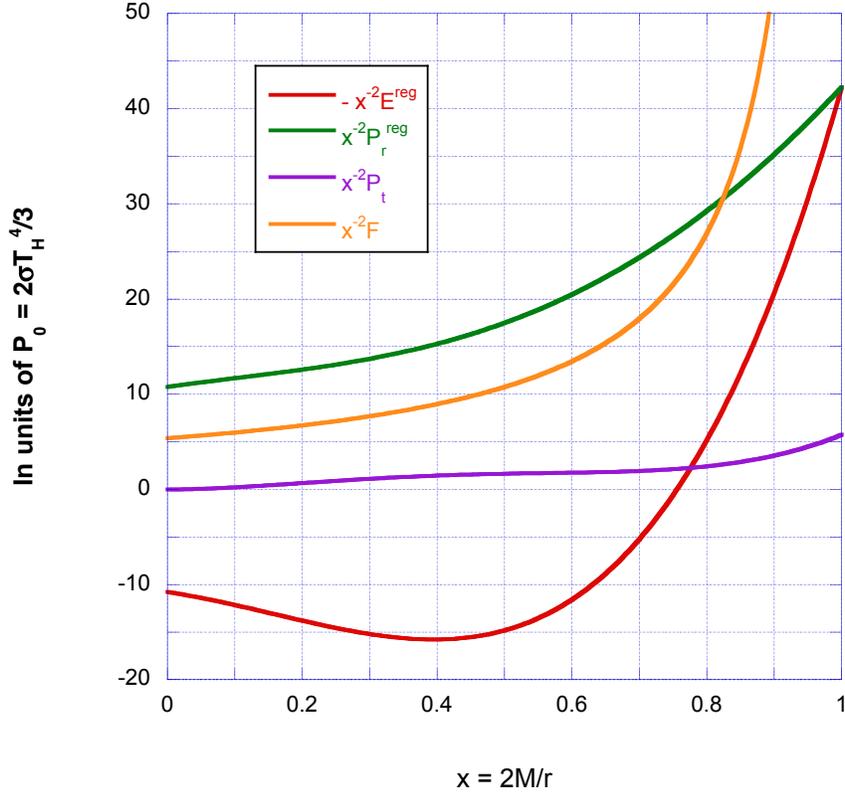

Figure 2. The static frame components of the "regular" part of spin 0 Unruh state SCSET are compared with the energy flux. Subtract $F$ from $E^{\text{reg}}$ and $P_r^{\text{reg}}$ to get the complete $E$ and $P_r$.

### d) Spin 1 Hartle-Hawking state

JO[12] found a rather good analytic approximation to the spin 1 HH $P_t$, and numerically calculated the correction to this analytic approximation. The published results were only presented as graphs, rather than as tables of data. However, a file of the numerical data for the spin 1 HH $P_r$ over the range $1 \leq r/2M \leq 3$ has been preserved by Visser[16] and is the basis of the fit discussed here.

The JO analytic approximation to the spin 1 HH state transverse stress is
$$P_t^A = 2P_0\left(1 + 2x + 3x^2 + 44x^3 - 305x^4 + 66x^5 - 579x^6\right), \qquad (2.22)$$
from which the corresponding radial stress is
$$P_r^A = 2P_0\left(1 + 2x + 3x^2 - 76x^3 + 295x^4 - 54x^5 + 285x^6\right). \qquad (2.23)$$
The difference $P_t - P_t^A \equiv \Delta_t$ was plotted in Fig. 4 of JO, but the scale of the graph was chosen to accommodate comparison with earlier very poor attempts at analytic approximations by Zel'nikov and Frolov[27] and by Brown, Ottewill and Page[28], making it very difficult to extract accurate numbers. I subtract $P_r^A$ from the $P_r$ Visser data file to get data for $\Delta_r$. A 3-6 polynomial fit, $\Delta_r = 2\sum_{n=3}^{6} d_n x^n$, is constrained by the closure condition $d_6 = (27/16)d_3 - (3/4)d_5$ and the condition $\Delta_r(1) = 0$ implied by $\Delta_t(1) = 0$ as verified numerically by JO. The result is

$$\Delta_r = 2P_0\left(-18.914x^3 + 122.111x^4 - 285.124x^5 + 181.927x^6\right), \quad (2.24)$$

with a reasonably good $\chi^2 = 0.29$. The corresponding $\Delta_t$ implied by momentum conservation gives

$$P_t = 2P_0\left(1 + 2x + 3x^2 + 25.086x^3 - 182.889x^4 - 219.124x^5 - 397.073x^6\right), \quad (2.25)$$

roughly the same as I found earlier relying on data from the JO graphs.

### d) Spin 1 Unruh state

The only published data on the transverse stress of the spin 1 Unruh state SCSET are graphs in JMO. Fig. 1 of JMO plots $P_t^U$ and their Fig. 4 plots the difference from the HH state $P_t^U - P_t^{HH}$ over the range $1 \leq r/2M \leq 2.5$. Neither are very satisfactory for extracting accurate numbers for $P_t^U$.

Recently I was given access to data files preserved by Visser[16] from the JMO spin 1 Unruh state calculations for the SCSET over the range $1 \leq r/2M \leq 3$. The graphs in JMO only extend to $r/2M = 2.5$. The data seems consistent with these graphs, within the limitations of the latter. However, there is no satisfactory polynomial fit to $x^{-3}P_t$ over the full range of the data for which the corresponding polynomial for $P_r^{reg}$ obtained from Eq. (2.9) has $r_2 = 2.4346$, as required for consistency with the asymptotic Hawking energy flux. I suspect the problem is poor accuracy of the JMO data at larger $r$, where $P_t$ is very small compared with its value at $x = 1$. and also very small compared to $P_t - P_t^{HH}$. A 3-6 fit to $P_t$ based only on data from $x > 0.4$ and constrained to give the correct $r_2$ is

$$P_t = 2P_0\left(40.90x^3 - 385.21x^4 + 32.69x^5 - 471.09x^6\right). \quad (2.26)$$

The residuals are less than 1% for $x > 0.5$, but increase to about 10% near $x = 0.4$. More accurate calculations extending to larger radii are necessary to have any confidence in the extrapolation to smaller values of $x$.

The $P_r^{reg}$ and $E^{reg}$ corresponding to Eq. (2.26) are

$$P_r^{reg} = 2P_0\left(2.435x^2 - 79.365x^3 + 367.20x^4 - 39.81x^5 + 220.25x^6\right) \quad (2.27)$$

and

$$E^{reg} = 2P_0\left(2.435x^2 + 2.435x^3 - 403.23x^4 + 25.57x^5 - 97.92x^6\right). \quad (2.28)$$

These are plotted, along with $P_t$ and the energy flux $F$, in Fig. 4. It is clear that the Hawking radiation is a very minor part of the spin 1 SCSET, in contrast to spin 0. For both spins $E^{reg}$ is negative in the vicinity of the horizon.

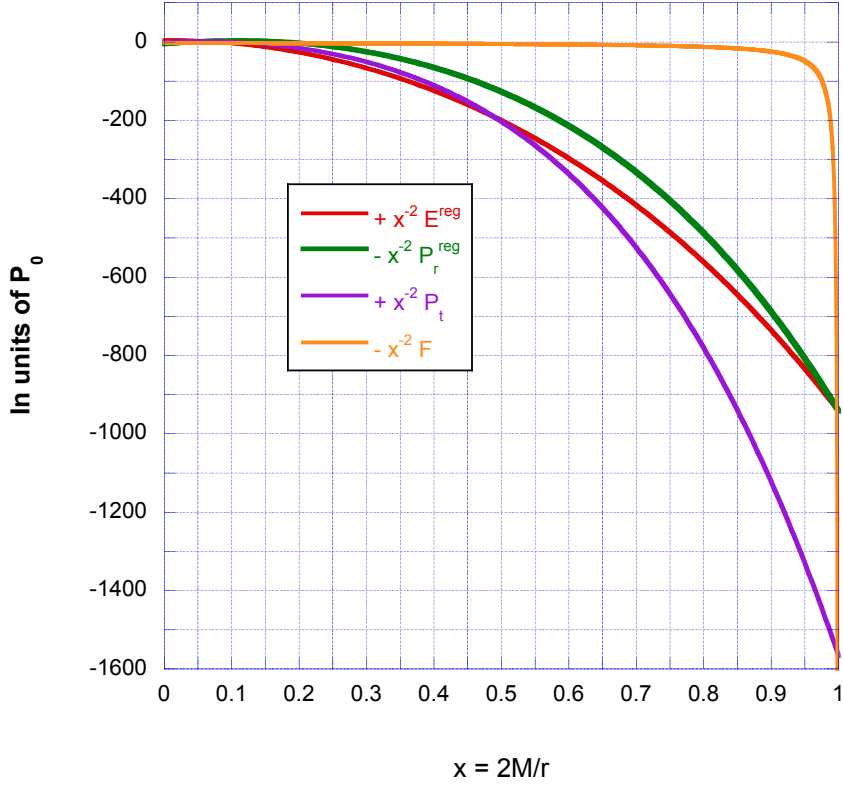

Figure 4. The various contributions to the spin 1 Unruh state SCSET are compared. Note the differences in signs from the similar plot for spin 0 in Fig. 2.

The spin 1 Hawking energy flux is insignificant in comparison with the other components of the SCSET, except *extremely* close to $x=1$ and near $x=0$. Vacuum polarization effects, as indicated by the trace anomaly, increase by an order of magnitude going from spin 0 to spin 1, while the Hawking luminosity decreases by more than a factor of two. At the horizon, the spin 1 transverse stress is $P_t = -1563 P_0$ and $E^{reg} = -P_r^{reg} = -939 P_0$, based on the Visser data file. Compare with the spin 0 results displayed in Fig. 2. That $E^{reg} < 0$ at the horizon for both spins is in spite of the difference in the sign of the trace anomaly.

e) Minimally coupled spin 0

Levi and Ori[29] have recently embarked on a program to greatly enhance the scope and accuracy of calculations of SCSETs for black holes. Their improved point-splitting renormalization technique[30] requires only one Killing vector field and no Wick rotation to a Euclidean metric. It can accommodate axisymmetric and nonstationary background geometries. Their initial results were for a *minimally-coupled* scalar field in the Schwarzschild background. All components of the SCSET were calculated directly out to $r = 50M$ ($x = 0.04$). They confirm the inflow of negative energy very close to the horizon and the gradual transition toward a purely radial outward flow of positive energy Hawking radiation at large radii. Their data are fit very well by 3-6 polynomials, with

$$P_t = P_0\left(1.524x^3 - 145.91x^4 + 108.19x^5 + 153.856x^6\right), \qquad (2.29)$$

$$T^\mu_\mu = P_0\left(2.22x^3 - 273.02x^4 + 365.55x^5 + 336.63x^6\right), \qquad (2.30)$$

$$P_r^{\text{reg}} = P_0\left(10.77x^2 + 7.72x^3 + 155.36x^4 - 17.17x^5 - 58.65x^6\right), \qquad (2.31)$$

and

$$E^{\text{reg}} = P_0\left(10.77x^2 + 8.55x^3 + 136.56x^4 - 166.34x^5 - 87.57x^6\right). \qquad (2.32)$$

The Hawking luminosity and $r_2$ are exactly the same, within numerical errors, as in the conformally coupled case. The trace and $P_t$, however, are much larger near the horizon.

### III. Physical interpretation of the SCSET

A positive energy flux, as in the Unruh state SCSET can be due to positive energy flowing out and/or negative energy flowing in. At large $r$ only positive energy outflow is physically acceptable. At the horizon only negative energy inflow in the static frame is consistent with a regular SCSET in a falling frame. If I had assumed outflow of positive energy just outside the horizon, as would have been appropriate if the Hawking radiation was generated within a Planck distance or so of the horizon, and therefore had defined the "regular" part of the SCSET by subtracting off an *outgoing* radial flow of *positive* energy, the coefficient $r_2$ in $P_r^{\text{reg}}$ would be zero, since then the outflow of positive energy at infinity would be completely assigned to the "singular" part of the SCSET. However, as I noted in Part II, this is inconsistent with the numerical results for the spin 0 transverse stress.

To get a better understanding of the transition from inflow at the horizon to outflow at large $r$, define an "outgoing" part of the net energy flux by

$$F^{\text{out}} \equiv \frac{1}{4}(E + P_r + 2F) = \frac{1}{4}\left(E^{\text{reg}} + P_r^{\text{reg}}\right) = \frac{1}{2}\left(P_r^{\text{reg}} + P_t^{\text{reg}}\right) - \frac{1}{4}T^\alpha_\alpha \equiv \frac{1}{2}(1-x)Z_s. \quad (3.1)$$

The "ingoing" part of the energy flux is

$$F^{\text{in}} \equiv F - F^{\text{out}} = \frac{1}{4}(-E - P_r + 2F). \qquad (3.2)$$

This strictly makes sense only if the SCSET is made up of radially propagating null "fluids" plus a radial-boost-invariant "vacuum polarization" contribution, but the ratio $F^{\text{out}}/F$ is a useful diagnostic if the energy flux is a major part of the SCSET. The quantity $Z_s$ is regular at the horizon and is also useful in making the Lorentz transformation from the static frame to a freely falling frame (see below). Evaluating $Z_0(x)$ using the conformally-coupled spin 0 Unruh state fit of Part II gives

$$Z_0 = P_0\left(10.77x^2 + 21.276x^3 + 31.914x^4 + 23.399x^5\right), \qquad (3.3)$$

and for the minimally coupled spin 0 fit

$$Z_0^{mc} = P_0 \left( 10.769 x^2 + 18.904 x^3 + 164.866 x^4 + 73.1097 x^5 \right). \tag{3.4}$$

The "outgoing" fraction of the net flux in the static frame is

$$F^{out} / F = (1-x)^2 Z_s / \left( h_s r_2 x^2 \right). \tag{3.5}$$

The outgoing fractions for conformally-coupled and minimally-coupled scalar fields are compared in Fig. 5. For conformal coupling the outgoing fraction monotonically decreases from one at $x = 0$ to zero at $x = 1$, with the transition from dominance of outgoing positive energy to dominance of ingoing negative energy at around $x = 2/3$ or $r = 3M$. For minimal coupling, the strictly radial flow interpretation would imply a mixture of outgoing positive energy and ingoing positive energy where $F^{out} / F > 1$, but this interpretation is probably not very appropriate, since the energy flux is rather small compared to the other components of the SCSET except very near the horizon and at rather large radii. In *neither* case are the results consistent with a physical picture in which the Hawking radiation, as has often been suggested in the literature, is due to pair creation or tunneling very close to the horizon.

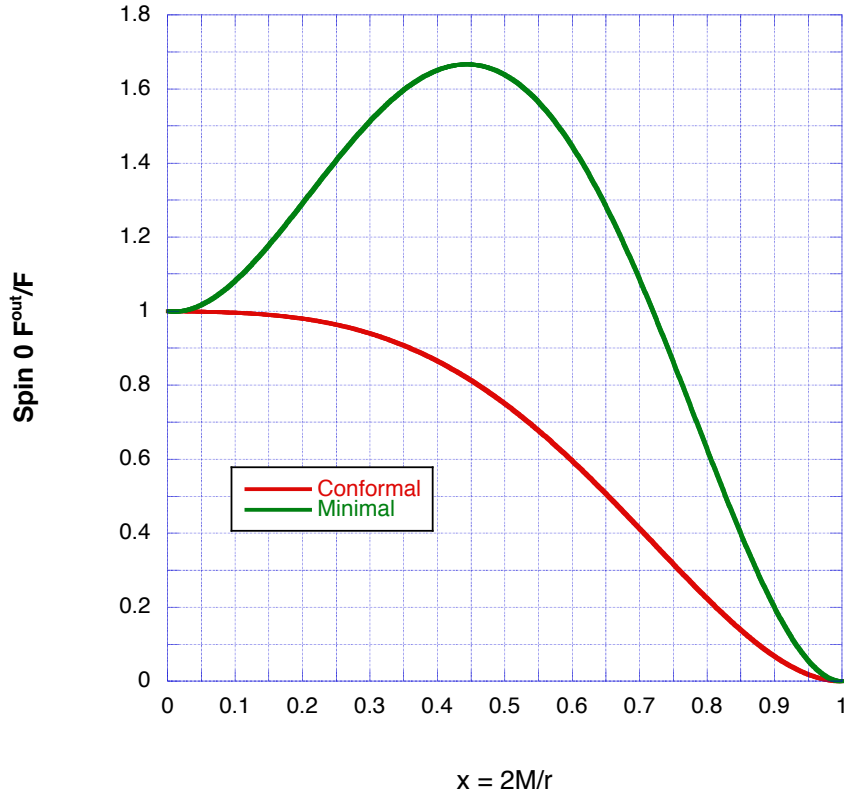

Figure 5. The "outgoing fraction" of the energy flux in the static frame is compared for conformally-coupled and minimally-coupled massless scalar fields in the Unruh state.

The lesson from the spin 0 SCSET is that as vacuum fluctuations propagating "outward" in the vicinity of the horizon are partially transmitted through and partially reflected from the potential barrier around $r = 3M$, they should be interpreted as physical particles only at larger radii, ultimately with respect to the Minkowski vacuum at future null infinity. The ingoing flow of negative energy across the horizon cannot be considered as associated with Hawking partner "particles", since physical particles should have positive energy. This is the same interpretation of the generation of Hawking radiation as originally proposed in papers by Unruh[31] and by Fulling[32]. Well inside the horizon it may or may not be appropriate to interpret the vacuum fluctuations as Hawking partner particles with locally positive energy and negative Killing energy, but in any case they are definitely entangled with the vacuum fluctuations that are interpreted as Hawking particles at large radii.

The situation is considerably more complicated for a spin 1 field. The energy flux is tiny compared with the other components of the SCSET except extremely close to the horizon, and the uncertainties in the numerical data and their extrapolation make isolating the outgoing part of the energy flux in any meaningful way impossible.

There are good physical reasons why pair creation or tunneling creating Hawking radiation extremely close to the horizon (as proposed, for instance, by Parikh and Wilczek[33]) doesn't make sense, besides being incompatible with the SCSET. A Hawking particle just outside the horizon would have to have an enormous energy and momentum in the static frame and an enormously larger energy and momentum still in a typical local free-fall frame in order to reach infinity with even the small energy corresponding to the Hawking temperature. The large local energy is *not* compensated by gravitational potential energy, since by the equivalence principle this has no local significance. The Hawking partner as a real particle in a local pair creation process must have a corresponding large energy and *outward* momentum in a local inertial frame straddling the horizon, in order for the partner to have negative Killing energy. This requires a doubly enormous violation of local conservation of energy and momentum.

The existence of Hawking radiation is due to the positive frequency vacuum modes at $\mathfrak{I}^-$ evolving into a mixture of positive and negative frequency modes at $\mathfrak{I}^+$, as argued eloquently by Hawking[1], but this says little about exactly where in the general vicinity of the black hole the particles are created, since the definition of a "particle" is highly ambiguous when their wavelengths are comparable to the curvature scale, as are the wavelengths of modes near the peak of the Hawking spectrum around $r = 3M$.

It is convenient to calculate the mismatch between the in and out vacuums by extrapolating the mode functions to the horizon of a stationary black hole. Visser[34] has given a nice pedagogical discussion based on Gullstrand-Painleve coordinates $(\tilde{t}, r)$ in Schwarzschild, which are regular on the horizon. In a WKB approximation, the phase of an outgoing scalar mode of frequency $\omega$ is $\mp i\left(\omega\tilde{t} - \int^r k(r')dr'\right)$. Near the horizon $(r - r_H \ll \kappa^{-1})$ the wave number $k \approx \omega / [\kappa(r - r_H) + i\varepsilon]$, with $\kappa$ the surface gravity of the horizon. The wave number diverges and changes sign at the horizon, since outgoing modes become ingoing in radius inside the horizon. The Feynman $i\varepsilon$ ensures the proper phase relation between a Hawking mode just outside the horizon and the "partner" mode just inside. However, any physical excitation is a wave packet, integrated over a range of frequencies. The rapid oscillation of phase near the horizon means that any such wave packet will have very small amplitude close to the horizon due to destructive interference between neighboring frequencies.

The short wavelength modes of an initial Minkowski vacuum state remain unexcited as long as they are localized very close to the horizon, where they are just propagating in the locally flat geometry. However, they are increasing redshifted relative to the uniformly accelerating static observers of the Schwarzschild spacetime, as they would be relative to Rindler observers in a flat spacetime. As their wavelengths become comparable to the radius of the black hole, they are disrupted by geodesic deviation (absent for a Rindler horizon). Far from the black hole it becomes possible to interpret the outgoing modes as physical particles relative to the asymptotic Minkowski vacuum. Modes contributing to the high frequency tail of the Hawking temperature Planck spectrum may never experience very much geodesic deviation, but they contribute very little to the asymptotic Hawking energy flux.

Is there any basis for the recent claim of Baker, et al[35] and others that there is significant entanglement of Hawking radiation with quantum fluctuations of the background geometry? Their argument is based on the huge energy of Hawking quanta created infinitesimally close to the horizon, but if Hawking quanta never have energy exceeding $\sim m_p^2 / M$ spread out over a distance of order $M$, the time delay induced by the backreaction as a fraction of the period of the Hawking wave is of order $(m_p^2 / M) M / M^2 \sim m_p^2 / M^2 \sim 10^{-76}$ for a solar mass black hole. The modes that give rise to Hawking radiation well after the formation of the black hole do have sub-Planckian wavelengths before the black hole is formed, but this is a Lorentz-frame-dependent statement.

Finally, I consider the transformation of the SCSET from the static frame to a frame freely falling from rest at infinity. The velocity of this frame with respect to the static frame is $v = -\sqrt{x}$. Making use of $Z_s(x)$ as defined in Eq. (3.1), the relevant physical components in the static frame can be written as

$$E = E^{\text{reg}} - F = 2(1-x)Z_s - P_r^{\text{reg}} - F, \quad P_r = P_r^{\text{reg}} - F. \tag{3.6}$$

The Lorentz transformation to the free fall frame is

$$E^{\text{ff}} = (1-x)^{-1}\left(E + xP_r + 2\sqrt{x}F\right), \quad P_r^{\text{ff}} = (1-x)^{-1}\left(xE + P_r + 2\sqrt{x}F\right),$$
$$F^{\text{ff}} = (1-x)^{-1}\left[\sqrt{x}(E + P_r) + (1+x)F\right]. \tag{3.7}$$

Then

$$E^{\text{ff}} = 2Z_s - P_r^{\text{reg}} - \frac{L_{\text{H}}}{4\pi r^2 \left(1+\sqrt{x}\right)^2}, \quad P_r^{\text{ff}} = 2xZ_s + P_r^{\text{reg}} - \frac{L_{\text{H}}}{4\pi r^2 \left(1+\sqrt{x}\right)^2},$$

$$F^{\text{ff}} = 2\sqrt{x}\, Z_s + \frac{L_{\text{H}}}{4\pi r^2 \left(1+\sqrt{x}\right)^2}.$$

(3.8)

The "outgoing" part of the energy flux in the free fall frame is

$$\left(F^{\text{ff}}\right)^{\text{out}} = \frac{1}{4}\left(E^{\text{ff}} + P_r^{\text{ff}} + 2F^{\text{ff}}\right) = \frac{1}{2}\left(1+\sqrt{x}\right)^2 Z_s, \tag{3.9}$$

and at the horizon is positive for spin 0, negative for spin 1. It is redshifted away to nothing in the static frame at the horizon. The "ingoing" part in the free fall frame is

$$\left(F^{\text{ff}}\right)^{\text{in}} = F^{\text{ff}} - \left(F^{\text{ff}}\right)^{\text{out}} = \frac{L_{\text{H}}}{4\pi r^2 \left(1+\sqrt{x}\right)^2} - \frac{1}{2}\left(1-\sqrt{x}\right)^2 Z_s. \tag{3.10}$$

## IV. SEMI-CLASSICAL BACKREACTION ON THE GEOMETRY

The expectation value of the renormalized SCSET can be inserted as a source in the classical Einstein equations to calculate first-order corrections to the classical spacetime geometry on which the calculation of the SCSET was based. This is not justified in all circumstances. One can imagine a "Schrodinger cat" quantum state that leads to a superposition of macroscopically different alternative geometries, rather than small fluctuations about a single classical history. However, in the evaporation of a large black hole, because a significant change in the geometry requires emission of an enormous number of Hawking quanta, and because the main effect is just a gradual decrease in the black hole's mass, semi-classical evolution makes sense initially. Approaching the Page time the backreaction is substantial, and it is no longer possible to consider the backreaction as a small perturbation of a single classical history[36]. The full quantum state then incorporates multiple classical histories, but new quantum fluctuations about each of them are still small, so it might be argued that the semi-classical backreaction is still approximately valid for the evolution along each of those classical histories.

To solve the Einstein equations with the SCSET source I work in advanced Eddington-Finkelstein coordinates $(v,r)$ with advanced time $v$ constant along ingoing radial null geodesics. A general spherically symmetric form of the metric in these coordinates is

$$ds^2 = -Ae^{2\psi}dv^2 + 2e^{\psi}dvdr + r^2\left(d\theta^2 + \sin^2\theta\, d\varphi^2\right), \tag{4.1}$$

following the notation of Bardeen[17]. The inverse metric has

$$g^{vv} = 0, \quad g^{vr} = e^{-\psi}, \quad g^{rr} = A \equiv 1 - \frac{2m}{r}. \tag{4.2}$$

The Einstein equations are then extremely simple, with

$$\left(\frac{\partial m}{\partial v}\right)_r = 4\pi r^2 T_v^r, \quad \left(\frac{\partial m}{\partial r}\right)_v = -4\pi r^2 T_v^v, \quad \left(\frac{\partial \psi}{\partial r}\right)_v = 4\pi r e^{\psi} T_r^v. \tag{4.3}$$

In terms of the components appearing in Eq. (4.3), $T_r^r \equiv T_v^v + Ae^\psi T_r^v$ and $T_\theta^\theta$ can be found from the momentum constraint equation $T_{r;\mu}^\mu = 0$.

The $v,r$ coordinate components of the SCSET in the Schwarzschild background, where $\psi = 0$ and $A = 1 - 2M/r$, can be written in terms of the static frame components as

$$T_v^v = -E - F = -E^{\text{reg}}, \tag{4.4}$$

$$T_r^r = P_r + F = P_r^{\text{reg}}, \quad T_\theta^\theta = T_\varphi^\varphi = P_t, \tag{4.5}$$

$$T_v^r = -(1-x)F = -\frac{3}{8}P_0 x^2 k_s, \tag{4.6}$$

$$T_r^v = (1-x)^{-1}(E + P_r + 2F) = (1-x)^{-1}(E^{\text{reg}} + P_r^{\text{reg}}) = 2Z_s. \tag{4.7}$$

All are perfectly finite and smooth at $x = 1$.

It follows immediately from Eq. (4.6) and the expression for $\partial m / \partial v$ in Eq. (4.3) that $\partial m / \partial v$ is the same at all radii at a given advanced time. With $E^{\text{reg}}$ falling off asymptotically as $(3/4)k_s P_0 x^2$, it would seem from the initial value equation for $\partial m / \partial r$ that $m$ should diverge linearly as $r \to \infty$, but as noted earlier this is an illusion. The asymptotic contribution to $m$ is just the energy of previously emitted Hawking radiation, but this is a finite amount of energy, since the black hole was formed at a finite time in the past. For the same reason, there is a cutoff to the logarithmic divergence in the radial integral for the metric function $\psi$. The geometry stays Schwarzschild to a very good approximation in the vicinity of the black hole, with corrections of order $m_p^2 / M^2$, but with a gradually decreasing gravitational mass.

There is an important missing piece of the semi-classical evolution of the black hole. We have not considered the part of the SCSET associated with quantum fluctuations of the gravitational field. The only available results are the spin 2 Hawking luminosity and the spin 2 trace anomaly. Since the spin 2 trace anomaly, with $q_2 = 212$, is more than an order of magnitude larger than the spin 1 trace anomaly, the quantum gravity contribution to the SCSET should overwhelmingly dominate that from lower spin fields, even though the spin 2 Hawking luminosity is much smaller than that of lower spins. Furthermore, general relativity is not conformally invariant, so there should be additional contributions to the trace of the SCSET not associated with the trace anomaly. It is even conceivable that the quantum gravity SCSET is qualitatively, as well as quantitatively, different from that of ordinary quantum fields.

## V. IMPLICATIONS FOR THE INFORMATION PARADOX

The black hole information paradox is usually stated as a conflict between the demands of quantum theory (unitary evolution with pure states evolving into pure states, monogamy of entanglement, and locality at least in the sense of causal propagation of quantum information), and the semi-classical evaporation of black holes. Hawking quanta are entangled with "Hawking partners" inside the black hole, and as the evaporation proceeds the black hole traps increasing amounts of quantum information, leaving the radiated Hawking quanta in a mixed state. Unless the trapped quantum information can somehow escape before the black hole evaporates completely, which would seem to require acausal propagation, the final result would apparently be evolution from an initial pure state to a final mixed state[18]. A Planck scale remnant containing the enormous amount of quantum information trapped in the evaporation of a large black hole is not an attractive prospect for a number of reasons. Unruh and Wald[37] have recently reiterated their quite compelling arguments as to why the standard picture of black hole evaporation should lead to a mixed quantum state in the exterior of the black hole, not a pure state, and argue that this is not a violation of fundamental principles of quantum field theory, but in fact is the natural, expected consequence of quantum field theory, even if the black hole evaporates completely without releasing its quantum information[38]. Still, there is a widely held belief, based in part on AdS-CFT, that there is a real paradox, and more or less exotic schemes have been proposed for how the trapped information may be able to escape[39].

Black hole complementarity[19] tried to argue that all quantum information on its way into the black hole is copied onto the event horizon and then gradually leaks out to infinity as subtle correlations in the apparently thermal Hawking emission, restoring a pure state for an external observer. The no-cloning theorem of quantum mechanics is not violated, it was claimed, because no single observer can detect both copies of the quantum information.[40] However, as long as quantum field theory and quantum gravity only allow a causal flow of quantum information on macroscopic scales I see no way that quantum information stored on the event horizon is a plausible solution to the black hole information problem. Rindler horizons are everywhere in Minkowski spacetime and certainly do not store quantum information. Also, the event horizon is an acausal construct that depends on on the entire future history of the black hole. No particular null hypersurface near the apparent horizon can be identified as the event horizon.

Almheiri, et al (AMPS)[41] have argued that substantial entanglement of the late Hawking radiation with the early Hawking radiation, by monogamy of entanglement, means that there cannot be the entanglement of Hawking particles with Hawking "partners" that makes it possible to sustain the standard semi-classical non-singular structure of the horizon as seen by a freely falling observer. This suggests that an observer freely falling across the horizon would be incinerated by a "firewall" of very high-energy excitations. The AMPS paper generated a firestorm of controversial proposals in the literature, which I will make no effort to discuss here. See reviews by Polchinski[42] and Marolf[43].

The energy conservation objections to pair creation very close to the horizon apply with even greater force to the creation of a firewall. Energetically the only way to have a firewall is to assume it is present at past null infinity, before the black hole has even formed. While there may be quantum states with this feature, these are not physically acceptable as states in our universe, as noted by Page.[44]

Hawking, et al[45] have suggested that a kind of quantum mechanical "soft hair" is associated with black hole event horizons, essentially zero energy photons and gravitons associated with an infinite degeneracy of the vacuum. Could this "soft hair" preserve the quantum information associated with accreting matter and the generation of Hawking radiation, which would eventually leak out as subtle correlations in the Hawking radiation? Mirbabayi and Porrati[46] and Bousso and Porrati[47] argue that this soft hair, as opposed to the soft hair at null infinity, is trivial and inherently incapable of carrying any quantum information. There is a sharing of the quantum information in the Hawking radiation between "hard" quanta and "soft" quanta, but quantum information is still lost inside the horizon of the black hole.

The Bekenstein-Hawking entropy $S_{BH} = A/(4m_p^2)$, where $A$ is the area of the horizon, $16\pi M^2$ for Schwarzschild, has an interpretation as the classical coarse-grained thermodynamic entropy of a black hole. See Wald[48] for a review. Is $S_{BH}$ a measure of the total number of quantum degrees of freedom of the black hole, as has been calculated in string theory for certain black holes with degenerate or nearly degenerate horizons[49]? If so, it should be an upper limit to the entanglement entropy. A recent argument to the contrary has been made by Rovelli[50].

I agree with Rovelli that a black hole is in general not a conventional quantum system with a fixed number of degrees of freedom proportional to its surface area. The vicinity of the horizon of a young black hole formed by stellar collapse is really just a rather empty region of spacetime. The physically appropriate microscopic measure of the entropy is the *entanglement* (von Neumann) entropy $S_{vN}$ of the black hole as a subsystem of the fields on a Cauchy hypersurface, renormalized so as *not* to include the short-range correlations of the vacuum across the horizon that are present across any sharp boundary. This is the total number of degrees of freedom in the interior entangled with the exterior of the black hole arising from the initial formation of the black hole, any subsequent accretion, and the entanglement generated by the emission of Hawking radiation. A black hole of mass $M$ *can* be formed by the collapse of $\sim S_{BH}$ entangled quanta with energy $\varepsilon_c \approx m_p^2/M$, but in the real world a young black hole's entanglement entropy $S_{vN}$ is typically tiny compared with $S_{BH}$, since a star whose collapse forms the black hole is made up of quanta with energies $\varepsilon \gg \varepsilon_c$. Emission of Hawking radiation causes $S_{vN}$ to increase, and in the absence of other influences it equals $S_{BH}$ at the Page time[51], when the black hole has lost about 1/2 of its original mass.

It is at the Page time that one is really forced to deal with the black hole information problem. There is nothing in the usual semi-classical theory of black hole evaporation that would explain why Hawking evaporation should stop at the Page time. If Hawking radiation continues to be emitted, and there is no way of retrieving quantum information from deep inside the black hole, the ratio $S_{vN}/S_{BH}$ would continue to increase and eventually become much larger than one.

This conclusion is based on the usual assumption of local quantum field theory, that there are vacuum modes with arbitrarily small wavelengths in an infinite-dimensional Hilbert space. What if, as suggested by holography and AdS-CFT, the Hilbert space is finite-dimensional, and the number of vacuum modes propagating along the horizon is finite and does not exceed the Bekenstein-Hawking entropy? Shouldn't the evolution of short wavelength vacuum modes into Hawking quanta then cease around the Page time, due to exhaustion of the initial supply of vacuum modes? After the Page time the black hole might become an inert massive remnant, still highly entangled with the earlier Hawking particles, but incapable of emitting more.

If this argument is correct, and there is renewed accretion of matter well after the Page time that substantially increases the mass of the black hole and the area of the event horizon, it is hard to see how the number of black hole microstates could increase to match the new, larger Bekenstein-Hawking entropy. The additional vacuum modes propagating along the enlarged horizon would have to come from a tiny sub-Planckian region just outside what would have been the event horizon if there had not been the late accretion. The number of quantum degrees of freedom associated with the accreted matter and radiation would typically be far from sufficient to make up the difference.

What is the fate of the quantum information trapped deep inside the black hole? Is it swallowed up by a singularity? The energy conditions required by the classical singularity theorems certainly can be violated locally in quantum field theory. Are there quantum versions of energy conditions that could be used to prove quantum singularity theorems? Or could quantum effects provide sufficient backreaction on the geometry to allow nonsingular evolution of the geometry inside the black hole, perhaps even ultimately removing all trapped surfaces and allowing the quantum information to escape? Could this happen by the Page time, so the entanglement entropy of the black hole never exceeds the Bekenstein-Hawking entropy?

One proposal for a quantum energy condition is the averaged null energy condition (ANEC),

$$\int T_{\alpha\beta} k^\alpha k^\beta \, d\lambda \geq 0, \tag{5.1}$$

where the integral extends over a complete null geodesic with tangent vector $k^\alpha = dx^\alpha / d\lambda$. It has been proven in a Minkowski spacetime and for the complete null geodesic generator of the horizon of a static black hole[52]. However, it is *not* true when evaluated with the spin 0 and spin 1 SCSETs of a Schwarzschild black hole, for null geodesics with nonzero angular momentum that have an inner turning point sufficiently close to $r = 3M$. It does seem to be valid when restricted to achronal null geodesics[53], no two points on which can be connected by a timelike curve. Fortunately, it can still be used with this restriction to prove significant results, like excluding traversable wormholes.

In an attempt to develop a quantum entropy bound, following a suggestion by Strominger and Thompson[54], Bousso, et al[20] define a generalized entropy $S_{\text{gen}}(\sigma)$ for a compact 2-surface $\sigma$ with area $A(\sigma)$ dividing a Cauchy hypersurface into two regions. With $S_{\text{out}}(\sigma)$ the von Neumann entropy of the "outer" non-compact region,

$$S_{\text{gen}}(\sigma) = S_{\text{out}}(\sigma) + \frac{A(\sigma)}{4m_{\text{p}}^2}. \tag{5.2}$$

This generalized entropy is used to formulate a "Quantum Focusing Conjecture" (QFC). Deforming $\sigma$ along null geodesics orthogonal to $\sigma$ gives compact 2-surfaces $\sigma'$. For $\sigma'$ that are uniform affine distance $\lambda$ from $\sigma$, a simplified version of the QFC states that $d^2 S_{\text{gen}}(\sigma')/d\lambda^2 \leq 0$, so if $dS_{\text{gen}}/d\lambda \leq 0$ initially, it remains non-positive until the null hypersurface (light sheet) hits a singularity or has a caustic. The QFC implies the "quantum Bousso bound"

$$S_{\text{out}}(\sigma') - S_{\text{out}}(\sigma) \leq \frac{[A(\sigma) - A(\sigma')]}{4 m_p^2}, \tag{5.3}$$

which in turn implies a quantum singularity theorem[55] analogous to the Penrose singularity theorem[20] in classical theory. The QFC would seem to imply no causal release of quantum information from the interior of a black hole without the presence of a Cauchy horizon or a singularity, and therefore a breakdown of unitarity.

The QFC does seem to be true in the context of the usual semi-classical theory of black hole evaporation. Consider the propagation of an "outward" radial null geodesic congruence from an initial two-sphere with radius $r_0$ just inside the apparent horizon of a large evaporating Schwarzschild black hole. Using the first-order semi-classical Einstein equations to relate derivatives of the metric functions in the geodesic equation to the SCSET gives

$$\frac{d^2 r}{d\lambda^2} = -4\pi r k^\alpha k^\beta T_{\alpha\beta} = -4\pi r e^\psi \left(\frac{dv}{d\lambda}\right)^2 \left[T_v^r + \left(\frac{dr}{dv}\right)^2 T_r^v\right] \tag{5.4}$$

in the coordinates of Eq. (4.1). For a Schwarzschild black hole with Hawking luminosity $L_H$, $T_v^r = -L_H/4\pi r^2$ dominates near the horizon, since $T_r^v$ is the same order as $T_v^r$ and $|dr/dv| \ll 1$ there. Consider a null congruence starting from $r = r_0$ at $v = 0$ with $(dr/dv)_0 \equiv \varepsilon \cong (r_0 - 2M)/4M$. The geodesic equation gives in the near horizon limit $dv/d\lambda \cong e^{-\kappa v}$, where $\kappa = 1/4M$ is the surface gravity of the horizon. Using this in Eq. (5.4) and solving for $r(v)$ gives

$$\frac{dr}{dv} = (\varepsilon + 2L_H) e^{\kappa v} - 2L_H. \tag{5.5}$$

Assuming only photons and gravitons contribute to the Hawking radiation, Page[51] has estimated that the rate of increase of the von Neumann entropy is

$$\frac{dS_{\text{vN}}}{dv} \approx \frac{1}{715 M}, \tag{5.6}$$

and in Eq. (1.2) for $L_H$, $\sum_s k_s = 7.23$. The rate of change of the generalized entropy near the horizon is then

$$\frac{dS_{\text{gen}}}{dv} = \frac{1}{715 M}\left[1 - 0{,}673 + 0.673\left(1 + \frac{\varepsilon}{2 L_H}\right)e^{\kappa v}\right]. \tag{5.7}$$

Any initial decrease of $S_{\text{gen}}$ requires $1 + \varepsilon/2 L_H < -0.485$, implying that the decrease accelerates to the future, consistent with the QFC. Of course, as an expectation value the SCSET at a given radius is really an average over the emission of many Hawking quanta, which requires a time interval $\Delta v \gg M$. Therefore, Eq. (5.7) should not be taken as a precise result.

What might conceivably lead to a different outcome, in violation of the QFC, is a deep connection between entanglement and geometry in quantum gravity generating a backreaction preventing the *renormalized* entanglement entropy across a compact 2-surface $\sigma$ from exceeding $A(\sigma)/4\hbar$. Some support for this idea comes from AdS-CFT, where Ryu and Takayanagi[56] have shown that entanglement in the CFT between disjoint parts of the AdS boundary has as a bulk dual an Einstein-Rosen bridge, with the entanglement entropy equal to 1/4 of its *minimal* area in Planck units. An explicit example was worked out by Jensen, et al[57]. This entropy bound is also consistent with a version of the ER=EPR conjecture of Maldacena and Susskind[58]. If microscopic Einstein-Rosen bridges connect entangled qbits, it is reasonable that there is a large backreaction modifying the macroscopic geometry when these approach a density of one per Planck area across a two-surface. The implications such an hypothesis was considered in a previous paper of mine[59]. However, a large backreaction on the macroscopic geometry where the spacetime curvature is highly sub-Planckian is quite a stretch from known physics.

An interesting alternative, motivated by the quantization of area in loop quantum gravity, is that the spacetime in the interior of the black hole can, when quantum backreaction becomes large, "tunnel" into a white hole without any singularity in the spacetime geometry. The quantum information that was trapped by the black hole reappears, after a long delay, in emission from the white hole. Unitarity of the quantum evolution demands the existence of an initial Cauchy hypersurface, i.e., the absence of any Cauchy horizon in the evolution of the black hole interior. Also, there must not be any naked singularity in the geometry where the evaporating black hole disappears and the white hole appears. A model for such a scenario has been presented recently by Bianchi, et al,[60] but is unsatisfactory in some ways. An improved version of their model, which has the flexibility to mesh with the results for the SCSET, is being prepared by Bardeen.[61]

## VI. SUMMARY

The semi-classical stress-energy tensors for conformally-coupled spin 0 and spin 1 fields in the exterior spacetime of a large Schwarzschild black hole were calculated in the 1980's and 1990's, but the results, over a limited range of radius, were not presented in a way that facilitated their physical interpretation. The attempts in this paper to find analytic fits to the numerical data that plausibly can be extrapolated to $r \gg 2M$ have been very successful for spin 0, but the accuracy of the calculations for the spin 1 Unruh state is not sufficient for a reliable extrapolation. The spin 1 calculations need to be redone and extended to larger radii. A project initiated by Levi and Ori[31] to revisit and extend the numerical results has the potential of clearing up these ambiguities as well as providing accurate results for Kerr and evolving black holes, though so far their results for the Schwarzschild background are only for minimally coupled, rather than conformally coupled, scalar fields.

For a large astrophysical black hole the semi-classical approximation is the first order of an expansion in powers of an incredibly small expansion parameter. There is no local significance for an event horizon, if one exists, and to challenge local quantum field theory in the vicinity of a large black hole would seem to also challenge the use of local quantum field theory in Minkowski spacetime as an approximation in gravitational fields of comparable strength in laboratories on the Earth, where it has been tested to exquisite precision. The semi-classical theory gives no indication that Hawking radiation is anything but a low energy phenomenon, associated with tidal disruption of vacuum fluctuations in the general vicinity of the black hole, and is completely inconsistent with Hawking radiation being generated by pair creation or tunneling within some small Planck scale neighborhood of the horizon.

My take on the black hole information problem is not new, but it has become unfashionable in certain quarters over the last 25 years or so. I see no way to prevent the great bulk of the quantum information associated with Hawking "partners" from ending up deep inside the black hole. As long as the black hole geometry near the horizon is close to Schwarzschild, the quantum information cannot be retrieved to purify the quantum fields outside the black hole without drastically acausal propagation, which would seem to be a much more serious violation of conventional quantum mechanics and quantum field theory than the failure to retrieve it. Furthermore, a Schwarzschild black hole horizon is locally indistinguishable from a Rindler horizon in Minkowski spacetime. Quantum information can cross a Rindler horizon without leaving behind any significant trace.

Even in the light of AdS-CFT correspondence, which implies that the bulk quantum fields must be unitary, I see no compelling reason why it is necessary for all the quantum information to escape to the AdS boundary. The quantum fields in the exterior of the black hole are a subsystem, which is not expected to be in a pure state even if the total system is in a pure state. Papadodimas and Raju[62] have suggested how the boundary CFT may be able to track the evolution of the black hole interior. This requires no acausal communication, since the quantum state evolves deterministically in both the bulk and the boundary.

Preventing a black hole from evaporating down to the Planck scale while retaining its trapped quantum information requires some sort of new physics. This new physics must kick in by the Page time, while the spacetime curvature in the vicinity of the horizon is still extremely sub-Planckian for a large black hole, if the Bekenstein-Hawking entropy is to be considered an upper limit to the entanglement entropy between the interior and the exterior of the black hole. Acausal propagation of quantum information should not be possible in regions of low curvature. Could the large macroscopic entanglement across the horizon distinguish the black hole horizon from what is locally just an ordinary null hypersurface in Minkowski spacetime, and generate sufficiently large departures from the Schwarzschild geometry to liberate trapped quantum information by the Page time? The QFC and its implications discussed in Part V would, if true, seem to guarantee the existence of a horizon as long as the black hole is large compared with the Planck scale. However, if the arguments of Bianchi, et al[60] regarding the quantum tunneling of a black hole to a white hole through what would have been the classical $r = 0$ singularity are correct, the quantum information trapped by the black hole will eventually reappear from the white hole, restoring purity of the quantum state.

ACKNOWLEDGEMENTS

I thank Andreas Karch and Ivan Muzinich for a number of helpful conversations, and Don Page for encouragement and help with the early literature. Also, I am very grateful to the Perimeter Institute for hosting a number of very stimulating visits that exposed me to a wide range of views on this and related topics, and during which some of the research for and writing of earlier versions of this paper was done. Research at Perimeter Institute is supported by the Government of Canada through the Department of Innovation, Science, and Economic Development, and by the Province of Ontario through the Ministry of Research and Innovation.